\documentclass[12pt,a4paper]{article}
\title{Noncommutative spacetime symmetries from covariant quantum mechanics}
\author{Alessandro Moia\\{\footnotesize\emph{Dipartimento di Fisica, Universit\`a ``La Sapienza'', P.le A. Moro 2, 00185 Roma, Italia}}}
\date{}
\usepackage[english]{babel}
\usepackage[latin1]{inputenc}
\usepackage{graphicx}
\usepackage{amsmath, amssymb}
\usepackage{fullpage}
\usepackage{units}
\usepackage{cite}

\newcommand{\act}{\triangleright}

\newcommand{\de}{\partial}

\begin{document}

\maketitle

\begin{abstract}
In the last decades, noncommutative spacetimes and their deformed relativistic symmetries have usually been studied in the context of field theory, replacing the ordinary Minkowski background with an algebra of noncommutative coordinates. However, spacetime noncommutativity can also be introduced into single-particle covariant quantum mechanics, replacing the commuting operators representing the particle's spacetime coordinates with noncommuting ones. In this paper, we provide a full characterization of a wide class of physically sensible single-particle noncommutative spacetime models and the associated deformed relativistic symmetries. In particular, we prove that they can all be obtained from the standard Minkowski model and the usual Poincar\'e transformations via a suitable change of variables. Contrary to previous studies, we find that spacetime noncommutativity does not affect the dispersion relation of a relativistic quantum particle, but only the transformation properties of its spacetime coordinates under translations and Lorentz transformations.
\end{abstract}

\section{Introduction}

In recent years, a sizeable literature has been devoted to exploring the physical consequences of assuming nontrivial commutation relations among spacetime coordinates. Much-studied examples include the so-called $\Theta$-Minkowski\cite{theta} and $\kappa$-Minkowski\cite{kappa} spacetimes, defined by
\begin{eqnarray}
\left[x^{\nu},x^{\lambda}\right]&=&i\ell^{2}\Theta^{\nu\lambda}\label{thm}
\end{eqnarray}
and
\begin{eqnarray}
\left[x^{\nu},x^{\lambda}\right]&=&i\ell(x^{\nu}\delta_{0}^{\ \lambda}-x^{\lambda}\delta_{0}^{\ \nu}),\label{km}
\end{eqnarray}
respectively. This research program is based on the idea that, regardless of the specific form a fully-fledged theory of quantum gravity may take, the quantization of the gravitational field should result in ordinary spacetime manifolds becoming some kind of noncommutative manifolds\cite{noncomm}. This possibility is particularly intriguing from a phenomenological point of view, because it could provide corrections to known physics before even addressing the problem of quantum gravity, by merely adapting existing physical theories to the new noncommutative framework. In particular, ordinary Poincar\'e symmetries of flat spacetime could be broken or deformed by noncommutativity, thereby leading to potentially observable quantum gravity effects even in the absence of strong gravitational fields\cite{quagrap}.

So far, most studies in this field have focused on developing suitable formal tools to deal with classical or quantum fields propagating on noncommutative backgrounds\cite{theta,GBR,chaichian,noeth}. In this context, ordinary Lie algebras have proved ill-suited to deal with deformed spacetime symmetries and have been replaced by more general structures called Hopf algebras\cite{hopf}. While this generalization makes sense and is indeed quite natural at the level of symmetry generators, it leads to puzzling results about infinitesimal symmetry transformations, such as the noncommutativity of transformation parameters and the impossibility of arbitrarily assigning their values\cite{noeth}.

In this paper, we advocate a completely different point of view about spacetime noncommutativity, pioneered to some extent in \cite{fuzzy1,fuzzy2}. We argue that noncommutative coordinates should be regarded as single-particle operators, like in standard, nonrelativistic quantum mechanics, rather than spacetime indices of classical or quantum fields. In this framework, symmetries are described as automorphisms of a canonical algebra, like in ordinary quantum mechanics, and there is no need for Hopf-algebraic concepts. We study generic spacetime noncommutativity of the form
\begin{eqnarray}
\left[x^{\nu},x^{\lambda}\right]=i\ell\Gamma^{\nu\lambda}_{\ \ \alpha}x^{\alpha}+i\ell^{2}\Theta^{\nu\lambda},
\end{eqnarray}
and provide a full characterization of the corresponding single-particle quantum models and the associated deformed relativistic symmetries. In particular, we prove that they can all be obtained from the commutative model and the standard Poincar\'e transformations by means of a suitable change of variables, thereby finding an explicit expression for the action of the deformed symmetries on the canonical variables. 

The paper is structured as follows. In Section 2, we discuss the physical import of spacetime noncommutativity and state our fundamental assumption. In Section 3, we quickly review single-particle covariant quantum mechanics\cite{RR}. In Sections 4 and 5, we introduce spacetime noncommutativity into this framework and find all possible deformed Poincar\'e symmetries which preserve the nontrivial commutation relations among the particle's coordinates. In Section 6, we discuss the general features of our models and compare our approach with the usual one based on Hopf algebras. In the last section we make some conclusive remarks.

\section{Which noncommutative coordinates?}

In theoretical physics, we can refer to two distinct concepts when talking about spacetime coordinates. On the one hand, we can mean arbitrary real functions defined on a spacetime manifold, like in differential geometry. These coordinates are just mathematical labels used to distinguish spacetime points and are not physical observables. A good example is given by spacetime coordinates in quantum field theory. In this context, observables are smeared field operators and coordinates only serve as a means of describing their relationships and tracing their dynamics. In fact, we can write quantum field theories with respect to arbitrary coordinate systems by changing variables in the equations of motion. The same can be said of classical general relativity, where the equations of motion are even covariant under general coordinate transformations. Let us call this first kind of coordinates `background coordinates'. On the other hand, we can refer to the observable spacetime position of some actual event with respect to some physical reference frame. In this case, coordinates are genuine dynamical quantities whose values can be theoretically computed and experimentally measured. A good example is given by the inertial cartesian spatial coordinates of a point particle at time $t_{0}$ in nonrelativistic quantum mechanics. It is clear that formal changes of variables can have no effects on such objects. Let us call this second kind of coordinates `particle coordinates'.

Schr\"odinger quantum field theory, i.e. second-quantized nonrelativistic quantum mechanics, provides us with an explicit expression of inertial particle spatial coordinates $x^{i}$ at time $t_{0}$ in terms of inertial background spatial coordinates $z^{i}$, thereby making manifest the deep conceptual difference between the two. Let a quantum field $\hat{\psi}(z,t)$ be a solution of the Schr\"odinger equation
\begin{eqnarray}
i\hbar\de_{t}\hat{\psi}(z,t)=\left(-\frac{\hbar^{2}}{2m}\nabla_{z}^{2}+V(z,t)\right)\hat{\psi}(z,t),
\end{eqnarray}
and let $\mathcal{H}^{1}$ be the Hilbert space of one-particle states $|1;\alpha\rangle$, defined by
\begin{eqnarray}
\hat{N}|1;\alpha\rangle=\int\hat{\psi}(z,t)^{\dagger}\hat{\psi}(z,t)d^{3}z|1;\alpha\rangle=|1;\alpha\rangle.
\end{eqnarray}
Then it is easy to verify that the observable
\begin{eqnarray}
\hat{x}^{i}(t_{0})=\int z^{i}\hat{\psi}(z,t_{0})^{\dagger}\hat{\psi}(z,t_{0})d^{3}z,
\end{eqnarray}
when restricted to $\mathcal{H}^{1}$, is the $i$-th particle coordinate at time $t_{0}$ of standard Heisenberg quantum mechanics. In this simple example, we clearly see that particle coordinates, being smeared field operators, are genuine observables, whereas background coordinates are mathematical labels devoid of direct physical meaning. In particular, we could write the theory in any background coordinate system and get back the same observable $\hat{x}^{i}(t_{0})$ by changing variables in the integral.

In the last decades, noncommutative spacetime coordinates have been proposed as an effective way of taking into account some quantum properties of the gravitational field without having to solve the full quantum gravity problem. The na\"ive idea is that, even when gravitational dynamics can be neglected and spacetime is Minkowski at large scales, the mere quantization of the gravitational degrees of freedom should result in spacetime points becoming fuzzy at scales of the order of the Planck length
\begin{eqnarray}
L_{P}=\sqrt{\frac{\hbar G}{c^{3}}},
\end{eqnarray}
in the same way as the mere quantization of single-particle degrees of freedom $q$ and $p$ results in phase space points becoming fuzzy at scales of the order of the Planck constant $h$. Drawing on this analogy, it is then natural to effectively model spacetime fuzziness through the introduction of nontrivial commutation relations among spacetime coordinates. In the light of our previous discussion, however, it is important to ask whether we are talking about background coordinates or particle coordinates. In order to make the vague suggestion of spacetime noncommutativity into a workable physical hypothesis, we must first of all give a definite answer to this question.

Most papers about noncommutative spacetime physics are based on the assumption of noncommutative background coordinates\cite{theta, GBR, chaichian, noeth}. In these studies, classical or quantum field theories are defined and characterized after replacing the ordinary Minkowski background with some noncommutative algebra of coordinates. Unlike nontrivial commutation properties of quantum observables, which reflect the incompatibility of the corresponding physical quantities, this kind of background noncommutativity does not admit a straightforward physical interpretation and appears therefore somewhat removed from the na\"ive intuition described above. Of course there are good reasons to explore this scenario, such as its actual relevance in 3D quantum gravity\cite{3D}, but we feel that the alternative point of view has not been sufficiently worked out for all its intuitive appeal and direct applicability to phenomenology.

In this paper, we will therefore assume that spacetime noncommutativity is a property of particle coordinates. The obvious analogy with nonrelativistic quantum mechanics and the well-established physical interpretation of noncommuting quantum observables make this, in our view, the most natural and straightforward assumption. Adopting this perspective, we will study the deformation of Poincar\'e symmetries induced by spacetime noncommutativity and eventually provide a full characterization of the resulting covariant quantum models.

\section{Covariant quantum mechanics}

In order to introduce spacetime noncommutativity into a single-particle quantum model, we must first of all address one fundamental problem. In the standard formulation of quantum mechanics, time is a classical evolution parameter. In particular, it is not an observable of the theory and it does not make mathematical nor physical sense to consider its commutation relations with the particle spatial coordinates. This serious obstruction undermined most attempts at modelling spacetime noncommutativity from a single-particle point of view and provided a strong reason for focusing on field theory and noncommutative background coordinates.

In the last decades, however, a covariant formulation of quantum mechanics has been developed which puts time and particle spatial coordinates on the same footing\cite{RR}. In this context, the extended phase space of the particle, including time $x^{0}$ and its conjugate momentum $-p_{0}$, is quantized and canonical coordinates $(p_{\mu},x^{\nu})$ become self-adjoint operators $(\hat{p}_{\mu},\hat{x}^{\nu})$ on a kinematical Hilbert space $\mathcal{K}$ satisfying canonical commutation relations
\begin{eqnarray}
\left[\hat{p}_{\mu},\hat{p}_{\tau}\right]&=&0,\label{ccs1}\\
\left[\hat{p}_{\mu},\hat{x}^{\nu}\right]&=&i\hbar\delta_{\mu}^{\ \nu},\\
\left[\hat{x}^{\nu},\hat{x}^{\lambda}\right]&=&0.\label{ccs2}
\end{eqnarray}
Starting from $\mathcal{K}$ and the extended canonical algebra $\mathcal{V}$ generated by $\hat{p}_{\mu}$ and $\hat{x}^{\nu}$, standard Heisenberg quantum mechanics is then recovered specifying a self-adjoint hamiltonian constraint $H(\hat{p}_{\alpha},\hat{x}^{\alpha})\in\mathcal{V}$. In particular, the physical Hilbert space $\mathcal{P}$ is given by the (improper) kernel of $H$ equipped with a suitably modified scalar product, while the usual algebra of Heisenberg observables is identified requiring that it commutes with $H$. The special constraint
\begin{eqnarray}
H_{nr}(\hat{p}_{\alpha},\hat{x}^{\alpha})=\hat{p}_{0}-H_{0}(\hat{p}^{k},\hat{x}^{k})
\end{eqnarray}
reproduces standard nonrelativistic quantum mechanics with hamiltonian $H_{0}$, but the covariant formalism is obviously more powerful. For example, the dynamics of a free relativistic scalar particle of mass $m$ is described by the quadratic constraint
\begin{eqnarray}
H_{r}(\hat{p}_{\alpha},\hat{x}^{\alpha})=\hat{p}^{\alpha}\hat{p}_{\alpha}-m^{2}c^{4}.\label{crel}
\end{eqnarray}

Since both time and particle spatial coordinates are represented by self-adjoint operators at the kinematical level, covariant quantum mechanics is the ideal tool for introducing arbitrary commutation relations among spacetime coordinates into a single-particle setting. It is sufficient to replace the trivial commutation rules (\ref{ccs2}) with more general ones. This realization, in the special case of $\kappa$-Minkowski spacetime, was the main ingredient of a pioneering but somewhat underappreciated work by Amelino-Camelia, Astuti and Rosati\cite{fuzzy1} and is in fact the starting point of the present paper. In the following, as anticipated in the introduction, we will study commutation rules of the form
\begin{eqnarray}
\left[\hat{x}^{\nu},\hat{x}^{\lambda}\right]=i\ell\Gamma^{\nu\lambda}_{\ \ \alpha}\hat{x}^{\alpha}+i\ell^{2}\Theta^{\nu\lambda},\label{mcr}
\end{eqnarray}
where $\ell$ is a fundamental length of the order of the Planck scale, while $\Gamma^{\nu\lambda}_{\ \ \alpha}$ and $\Theta^{\nu\lambda}$ are constant dimensionless matrices antisymmetric in the indices $\nu$ and $\lambda$. These are the most general commutation relations which trivialize in the limit $\ell\rightarrow0$ and are analytic in both the spacetime coordinates $\hat{x}^{\nu}$ and the deformation parameter $\ell$. They include the popular $\Theta$-Minkowski (\ref{thm}) and $\kappa$-Minkowski (\ref{km}) as particular cases.

\section{Deformed canonical algebra}

First of all, we must redefine the extended canonical algebra in the noncommutative models. Taking the canonical commutation relations (\ref{ccs1})-(\ref{ccs2}) and replacing (\ref{ccs2}) with (\ref{mcr}), we obtain
\begin{eqnarray}
\left[\hat{p}_{\mu},\hat{p}_{\tau}\right]&=&0,\\
\left[\hat{p}_{\mu},\hat{x}^{\nu}\right]&=&i\hbar\delta_{\mu}^{\ \nu},\label{Heis}\\
\left[\hat{x}^{\nu},\hat{x}^{\lambda}\right]&=&i\ell\Gamma^{\nu\lambda}_{\ \ \alpha}\hat{x}^{\alpha}+i\ell^{2}\Theta^{\nu\lambda}.
\end{eqnarray}
These commutation rules are not consistent in general, because the Jacobi identities for $\hat{p}_{\mu}$, $\hat{x}^{\nu}$ and $\hat{x}^{\lambda}$ are violated. To take care of this problem, we allow for a momentum-dependent deformation of the Heisenberg relations (\ref{Heis}) and write
\begin{eqnarray}
\left[\hat{p}_{\mu},\hat{p}_{\tau}\right]&=&0,\label{ccm1}\\
\left[\hat{p}_{\mu},\hat{x}^{\nu}\right]&=&i\hbar[\Delta(\ell\hat{p})]_{\mu}^{\ \nu},\\
\left[\hat{x}^{\nu},\hat{x}^{\lambda}\right]&=&i\ell\Gamma^{\nu\lambda}_{\ \ \alpha}\hat{x}^{\alpha}+i\ell^{2}\Theta^{\nu\lambda}.\label{ccm2}
\end{eqnarray}
A priori, the dimensionless matrix $\Delta_{\mu}^{\ \nu}$ could also depend on $\ell^{-1}\hat{x}^{\nu}$, but we are ruling out this possibility in order to avoid non-analyticity in either the spacetime coordinates $\hat{x}^{\nu}$ or the deformation parameter $\ell$. We are also leaving (\ref{ccm1}) undeformed because we are assuming that gravity is negligible and spacetime is flat. In order for (\ref{ccm1})-(\ref{ccm2}) to be consistent and reduce to (\ref{ccs1})-(\ref{ccs2}) in the commutative limit, $\Delta_{\mu}^{\ \nu}$ must satisfy $[\Delta(0)]_{\mu}^{\ \nu}=\delta_{\mu}^{\ \nu}$ and
\begin{eqnarray}
\ell\Gamma^{\nu\lambda}_{\ \ \alpha}\Delta_{\mu}^{\ \alpha}+\hbar\Delta_{\alpha}^{\ \nu}\de_{\alpha}\Delta_{\mu}^{\ \lambda}-\hbar\Delta_{\alpha}^{\ \lambda}\de_{\alpha}\Delta_{\mu}^{\ \nu}=0,\label{delta}
\end{eqnarray}
where we have set $\de_{\alpha}=\de/\de p_{\alpha}$. Since these conditions do not determine $\Delta_{\mu}^{\ \nu}$ uniquely, we must conclude that the modified commutation relations (\ref{mcr}) are not sufficient to fully characterize spacetime noncommutativity in our covariant single-particle setting, but must be complemented by a compatible deformation of the Heisenberg relations. It is worth explicitly pointing out that different choices of $\Delta_{\mu}^{\ \nu}$ are not physically equivalent, because they determine different uncertainty relations between particle coordinates and momenta.

Assuming that $\Delta_{\mu}^{\ \nu}$ is an invertible matrix, we can define a set of deformed self-adjoint coordinates $\hat{q}^{\nu}$ as
\begin{eqnarray}
\hat{q}^{\nu}&=&\frac{1}{2}\left\lbrace[\hat{x}^{\alpha}-\ell\Sigma^{\alpha}(\ell\hat{p})][\Delta^{-1}(\ell\hat{p})]_{\alpha}^{\ \nu}+\mathrm{h.c.}\right\rbrace=\nonumber\\
&=&\hat{x}^{\alpha}[\Delta^{-1}(\ell\hat{p})]_{\alpha}^{\ \nu}+\frac{1}{2}i\hbar[\Delta(\ell\hat{p})]_{\gamma}^{\ \alpha}\de_{\gamma}[\Delta^{-1}(\ell\hat{p})]_{\alpha}^{\ \nu}-\ell\Sigma^{\alpha}(\ell\hat{p})[\Delta^{-1}(\ell\hat{p})]_{\alpha}^{\ \nu}=\nonumber\\
&=&[\hat{x}^{\alpha}-\ell\Sigma^{\alpha}(\ell\hat{p})-i\ell\Omega^{\alpha}(\ell\hat{p})][\Delta^{-1}(\ell\hat{p})]_{\alpha}^{\ \nu},\label{q}
\end{eqnarray}
where $\Sigma^{\alpha}$ is a still unspecified vector depending on the momenta and we have introduced the shorthand notation
\begin{eqnarray}
\Omega^{\alpha}(\ell\hat{p})=\frac{\hbar}{2\ell}\de_{\gamma}[\Delta(\ell\hat{p})]_{\gamma}^{\ \alpha}.
\end{eqnarray}
Inverting the previous relations, we can express $\hat{x}^{\nu}$ as functions of $\hat{q}^{\nu}$ and $\hat{p}_{\mu}$:
\begin{eqnarray}
\hat{x}^{\nu}=\frac{1}{2}\left\lbrace\hat{q}^{\alpha}[\Delta(\ell\hat{p})]_{\alpha}^{\ \nu}+\ell\Sigma^{\nu}(\ell\hat{p})+\mathrm{h.c.}\right\rbrace=\hat{q}^{\alpha}[\Delta(\ell\hat{p})]_{\alpha}^{\ \nu}+\ell\Sigma^{\nu}(\ell\hat{p})+i\ell\Omega^{\nu}(\ell\hat{p}).\label{x}
\end{eqnarray}
This change of variables is useful because we can make $\hat{p}_{\mu}$ and $\hat{q}^{\nu}$ satisfy canonical commutation relations by appropriately choosing $\Sigma^{\alpha}$. In fact, computing the relevant commutators, we find
\begin{eqnarray}
\left[\hat{p}_{\mu},\hat{q}^{\nu}\right]=i\hbar\delta_{\mu}^{\ \nu},
\end{eqnarray}
irrespectively of $\Sigma^{\alpha}$, and
\begin{eqnarray}
\left[\hat{q}^{\nu},\hat{q}^{\lambda}\right]&=&\frac{1}{2}\left\lbrace(\Delta^{-1})_{\alpha}^{\ \nu}(\Delta^{-1})_{\beta}^{\ \lambda}\left[(i\ell\Gamma^{\alpha\beta}_{\ \ \gamma}\Delta_{\mu}^{\ \gamma}+i\hbar\Delta_{\gamma}^{\ \alpha}\de_{\gamma}\Delta_{\mu}^{\ \beta}-i\hbar\Delta_{\gamma}^{\ \beta}\de_{\gamma}\Delta_{\mu}^{\ \alpha})\hat{q}^{\mu}+\right.\right.\nonumber\\
&&\left.\left.\mbox{}+(i\hbar\ell\Delta_{\gamma}^{\ \alpha}\de_{\gamma}\Sigma^{\beta}-i\hbar\ell\Delta_{\gamma}^{\ \beta}\de_{\gamma}\Sigma^{\alpha}+i\ell^{2}\Gamma^{\alpha\beta}_{\ \ \gamma}\Sigma^{\gamma}+i\ell^{2}\Theta^{\alpha\beta})\right]+\mathrm{h.c.}\right\rbrace.
\end{eqnarray}
The first term in square brackets vanishes because of the identities (\ref{delta}), and the second can be put to zero by choosing $\Sigma^{\alpha}$ such that
\begin{eqnarray}
i\hbar\Delta_{\gamma}^{\ \alpha}\de_{\gamma}\Sigma^{\beta}-i\hbar\Delta_{\gamma}^{\ \beta}\de_{\gamma}\Sigma^{\alpha}+i\ell\Gamma^{\alpha\beta}_{\ \ \gamma}\Sigma^{\gamma}+i\ell\Theta^{\alpha\beta}=0.\label{sigma}
\end{eqnarray}
Therefore, we can describe our deformed canonical algebra $(\hat{p}_{\mu},\hat{x}^{\nu})$ as just a standard canonical algebra $(\hat{p}_{\mu},\hat{q}^{\nu})$ equipped with momentum-dependent functions $\Delta_{\mu}^{\ \nu}$ and $\Sigma^{\alpha}$ satisfying (\ref{delta}) and (\ref{sigma}), respectively. Noncommutative coordinates $\hat{x}^{\nu}$ are then given by (\ref{x}). This description of $(\hat{p}_{\mu},\hat{x}^{\nu})$ is a generalization of the concept of pregeometric representation introduced and developed in \cite{GBR} and \cite{fuzzy1}, respectively. Since conditions (\ref{sigma}) are not sufficient to univocally determine $\Sigma^{\alpha}$, different choices of $\Sigma^{\alpha}$ are associated with different, physically equivalent representations of the deformed canonical algebra. Commutative coordinates $\hat{q}_{\Sigma}^{\nu}$ and $\hat{q}_{\Sigma+\delta\Sigma}^{\nu}$ corresponding to representations $\Sigma^{\alpha}$ and $\Sigma^{\alpha}+\delta\Sigma^{\alpha}$ via (\ref{q}) are related by
\begin{eqnarray}
\hat{q}^{\nu}_{\Sigma+\delta\Sigma}=\hat{q}^{\nu}_{\Sigma}-\ell\delta\Sigma^{\alpha}[\Delta^{-1}]_{\alpha}^{\ \nu}.
\end{eqnarray}

\section{Deformed Poincar\'e symmetries}

We are now ready to discuss deformed relativistic symmetries in our single-particle framework. In covariant quantum mechanics, continuous groups of kinematical symmetries are described by continuous groups of automorphisms of the extended canonical algebra $\mathcal{V}$. Such transformations can always be unitarily implemented on $\mathcal{K}$ and are fully characterized by a set of self-adjoint generators $\hat{g}_{i}\in\mathcal{V}$. Kinematical symmetries which leave invariant the hamiltonian constraint $H$ are automatically automorphisms of the algebra of Heisenberg observables, and can therefore be identified as actual physical symmetries like in ordinary quantum mechanics. A Poincar\'e transformation $(\Lambda,a)$, for example, is described by the following map:
\begin{eqnarray}
\hat{x}^{\nu}&\mapsto&\Lambda^{\nu}_{\ \alpha}\hat{x}^{\alpha}+a^{\nu},\label{poin1}\\
\hat{p}_{\mu}&\mapsto&\Lambda_{\mu}^{\ \beta}\hat{p}_{\beta},\label{poin2}
\end{eqnarray}
which obviously preserves canonical commutation relations (\ref{ccs1})-(\ref{ccs2}). This group of automorphisms is generated by the self-adjoint operators $\hat{p}_{\mu}$ and $\hat{m}_{\rho\sigma}$, where
\begin{eqnarray}
\hat{m}_{\rho\sigma}=\hat{x}_{\rho}\hat{p}_{\sigma}-\hat{x}_{\sigma}\hat{p}_{\rho}.
\end{eqnarray}
If we impose the Poincar\'e-invariant constraint $H_{r}$ given in (\ref{crel}), $\hat{p}_{\mu}$ and $\hat{m}_{\rho\sigma}$ become constants of motion and the resulting model, describing a free relativistic scalar particle, is Poincar\'e-symmetric.

When we introduce spacetime noncommutativity, Poincar\'e transformations are not kinematical symmetries anymore. In fact, the modified commutation rules (\ref{mcr}) are not preserved in general by the action (\ref{poin1})-(\ref{poin2}). There are two alternative attitudes we can take towards this breaking of standard relativistic symmetries. On the one hand, we can view it as evidence of the failure of the relativity principle and the existence of preferred reference frames. On the other hand, we can just take it as an indication that ordinary Poincar\'e transformations are inadequate to describe relativistic symmetries in this regime and must be deformed to accomodate the new fundamental scale $\ell$, in the same way as Galileo transformations had to be deformed to accomodate the universal speed constant $c$. We adopt this second perspective, usually referred to as DSR in the literature\cite{DSR}, and assume that our noncommutative models admit deformed relativistic symmetries which reduce to Poincar\'e transformations in the limit $\ell\rightarrow0$.

A priori, the kinematical symmetry group could be discontinuous in the presence of spacetime noncommutativity. If this were the case, though, we could not even speak of symmetry generators and the commutative limit would be exceedingly singular to be dealt with. Therefore, we will rule out this possibility and assume that deformed relativistic symmetries are described by a 10-dimensional Lie group of automorphisms of the deformed canonical algebra (\ref{ccm1})-(\ref{ccm2}), like in the commutative case. Even requiring that these transformations reduce to standard Poincar\'e symmetries in the limit $\ell\rightarrow0$, the problem is obviously underconstrained and we expect to find many possible deformations of the usual relativistic symmetries for any given noncommutative single-particle model (\ref{ccm1})-(\ref{ccm2}). Our aim is to characterize as sharply as we can these possibilities.

First of all, we observe that deformed translations must be generated by $\hat{p}_{\mu}$, because momenta are physically defined as the generators of spacetime translations. Let us then denote the deformed Lorentz generators with $\hat{m}_{\rho\sigma}$, like in the commutative case. The deformed symmetry algebra generated by ($\hat{p}_{\mu}$,$\hat{m}_{\rho\sigma}$) must contract to the Poincar\'e algebra in the limit $\ell\rightarrow0$. However, it is a well-known result by Levy-Nahas\cite{LN} that the only Lie algebra deformations of the Poincar\'e algebra are the de Sitter and anti de Sitter algebras. Since we are assuming that spacetime is flat, we can conclude that $\hat{p}_{\mu}$ and $\hat{m}_{\rho\sigma}$ must satisfy their usual commutation relations, i.e.
\begin{eqnarray}
\left[\hat{p}_{\mu},\hat{p}_{\tau}\right]&=&0,\\
\left[\hat{p}_{\mu},\hat{m}_{\rho\sigma}\right]&=&i\hbar(g_{\rho\mu}\hat{p}_{\sigma}-g_{\mu\sigma}\hat{p}_{\rho}),\label{pm}\\
\left[\hat{m}_{\mu\nu},\hat{m}_{\rho\sigma}\right]&=&i\hbar\left(g_{\rho\nu}\hat{m}_{\mu\sigma}-g_{\mu\rho}\hat{m}_{\nu\sigma}+g_{\sigma\nu}\hat{m}_{\rho\mu}-g_{\mu\sigma}\hat{m}_{\rho\nu}\right),\label{mm}
\end{eqnarray}
even in the noncommutative case. This means that deformed Poincar\'e symmetries have their usual action (\ref{poin2}) on momenta, with the deformation only affecting the transformation (\ref{poin1}) of spacetime coordinates.

To complete our analysis, it is convenient to choose a representation $\Sigma^{\alpha}$ and express $\hat{m}_{\rho\sigma}$ as functions of the canonical variables $(\hat{p}_{\mu},\hat{q}^{\nu}_{\Sigma})$. It now follows from (\ref{pm}) that $\hat{m}_{\rho\sigma}$ must be linear in $\hat{q}^{\nu}_{\Sigma}$, so that we can generically write
\begin{eqnarray}
\hat{m}_{\rho\sigma}=\left[\hat{q}_{\rho}^{\Sigma}+\ell\Phi_{\rho}(\ell\hat{p})\right]\hat{p}_{\sigma}-\left[\hat{q}_{\sigma}^{\Sigma}+\ell\Phi_{\sigma}(\ell\hat{p})\right]\hat{p}_{\rho}.
\end{eqnarray}
Requiring that $\hat{m}_{\rho\sigma}$ satisfy the last commutation rules (\ref{mm}), we obtain the following conditions on $\Phi_{\rho}$:
\begin{eqnarray}
g_{\sigma\gamma}\de_{\gamma}\Phi_{\rho}-g_{\rho\gamma}\de_{\gamma}\Phi_{\sigma}=0.\label{condphi}
\end{eqnarray}
If we make the substitution
\begin{eqnarray}
\Phi^{\rho}=\delta\Sigma^{\alpha}(\Delta^{-1})_{\alpha}^{\ \rho},
\end{eqnarray}
a tedious but straightforward calculation shows that $\Phi_{\rho}$ satisfy (\ref{condphi}) if and only if $\delta\Sigma^{\alpha}$ satisfy the homogeneous version of (\ref{sigma}). As a consequence, $\overline{\Sigma}^{\alpha}=\Sigma^{\alpha}-\delta\Sigma^{\alpha}$ defines another representation of the deformed canonical algebra. Writing $\hat{m}_{\rho\sigma}$ in terms of the canonical variables $(\hat{p}_{\mu},\hat{q}^{\nu}_{\overline{\Sigma}})$, we obtain at last
\begin{eqnarray}
\hat{m}_{\rho\sigma}=\hat{q}_{\rho}^{\overline{\Sigma}}\hat{p}_{\sigma}-\hat{q}_{\sigma}^{\overline{\Sigma}}\hat{p}_{\rho}.
\end{eqnarray}
In other words, we have proved that it is always possible to find a unique set of commutative coordinates $\hat{q}^{\nu}_{\overline{\Sigma}}$ which transform like standard 4-vectors under the action of the deformed Lorentz symmetries. This means that the corresponding representation $\overline{\Sigma}^{\alpha}$ univocally determines the action of the deformed Lorentz transformations on the deformed canonical algebra and is therefore physically distinguished from the others.

We can finally provide a complete and very compact characterization of all possible single-particle quantum models of spacetime noncommutativity (\ref{mcr}) and their deformed relativistic symmetries. They are all obtained from a standard canonical algebra $(\hat{p}_{\mu},\hat{q}^{\nu})$ by specifying an invertible matrix $[\Delta(\ell\hat{p})]_{\mu}^{\ \nu}$ and a vector $\Sigma^{\alpha}(\ell\hat{p})$ satisfying (\ref{delta}), (\ref{sigma}) and the boundary conditions $[\Delta(0)]_{\mu}^{\ \nu}=\delta_{\mu}^{\ \nu}$. Noncommutative spacetime coordinates $\hat{x}^{\nu}$ are defined via (\ref{x}) and the action of deformed relativistic symmetries $(\Lambda,a)$ is given by the ordinary Poincar\'e action on the standard canonical coordinates $(\hat{p}_{\mu},\hat{q}^{\nu})$:
\begin{eqnarray}
\hat{q}^{\nu}&\mapsto&\Lambda^{\nu}_{\ \alpha}\hat{q}^{\alpha}+a^{\nu},\\
\hat{p}_{\mu}&\mapsto&\Lambda_{\mu}^{\ \beta}\hat{p}_{\beta}.
\end{eqnarray}
This results in a deformed action on the spacetime coordinates $\hat{x}^{\nu}$, given by
\begin{eqnarray}
\hat{x}^{\nu}&\mapsto&\hat{x}^{\gamma}[\Delta^{-1}(\ell\hat{p})]_{\gamma}^{\ \beta}\Lambda^{\alpha}_{\ \beta}[\Delta(\ell\Lambda\hat{p})]_{\alpha}^{\ \nu}+a^{\alpha}[\Delta(\ell\Lambda\hat{p})]_{\alpha}^{\ \nu}+\ell\left[\Sigma^{\nu}(\ell\Lambda\hat{p})+i\Omega^{\nu}(\ell\Lambda\hat{p})\right]-\nonumber\\
&&\mbox{}-\ell\left[\Sigma^{\gamma}(\ell\hat{p})+i\Omega^{\gamma}(\ell\hat{p})\right][\Delta^{-1}(\ell\hat{p})]_{\gamma}^{\ \beta}\Lambda^{\alpha}_{\ \beta}[\Delta(\ell\Lambda\hat{p})]_{\alpha}^{\ \nu}.
\end{eqnarray}
The deformed symmetry group is generated by the momenta $\hat{p}_{\mu}$ and the self-adjoint operators
\begin{eqnarray}
\hat{m}_{\rho\sigma}=\hat{q}_{\rho}\hat{p}_{\sigma}-\hat{q}_{\sigma}\hat{p}_{\rho},
\end{eqnarray}
and the corresponding infinitesimal variations of the coordinates $\hat{x}^{\nu}$ are given by
\begin{eqnarray}
\delta_{\varepsilon}\hat{x}^{\nu}&=&\frac{1}{i\hbar}\varepsilon^{\mu}[\hat{p}_{\mu},\hat{x}^{\nu}]=\varepsilon^{\mu}[\Delta(\ell\hat{p})]_{\mu}^{\ \nu},\\
\delta_{\varphi}\hat{x}^{\nu}&=&\frac{1}{i\hbar}\varphi^{\rho\sigma}[\hat{m}_{\rho\sigma},\hat{x}^{\nu}]=\nonumber\\
&=&2\varphi^{\rho\sigma}\left\lbrace\hat{x}^{\gamma}[\Delta^{-1}(\ell\hat{p})]_{\gamma}^{\ \alpha}\left(g_{\alpha[\rho}[\Delta(\ell\hat{p})]_{\sigma]}^{\ \nu}+\hat{p}_{[\rho}\de^{\sigma]}[\Delta(\ell\hat{p})]_{\alpha}^{\ \nu}\right)-\right.\nonumber\\
&&\mbox{}-\ell\left[\Sigma^{\gamma}(\ell\hat{p})+i\Omega^{\gamma}(\ell\hat{p})\right][\Delta^{-1}(\ell\hat{p})]_{\gamma}^{\ \alpha}\left(g_{\alpha[\rho}[\Delta(\ell\hat{p})]_{\sigma]}^{\ \nu}+\hat{p}_{[\rho}\de^{\sigma]}[\Delta(\ell\hat{p})]_{\alpha}^{\ \nu}\right)+\nonumber\\
&&\left.\mbox{}+\ell\hat{p}_{[\rho}\de^{\sigma]}\left[\Sigma^{\nu}(\ell\hat{p})+i\Omega^{\nu}(\ell\hat{p})\right]\right\rbrace,
\end{eqnarray}
where little square brackets denote antisymmetrization.

In order to obtain a complete covariant quantum model, we must still specify a hamiltonian constraint which is invariant under deformed Poincar\'e symmetries and reduces to the usual one in the commutative limit. In the light of our previous findings, however, the problem is trivial. In fact, since momenta have their usual transformation properties, the undeformed relativistic contraint (\ref{crel}) is invariant under deformed symmetries and is therefore the only natural choice.

\section{Discussion and comparison with other approaches}

Having obtained a precise characterization of all possible single-particle quantum models of spacetime noncommutativity (\ref{mcr}), we can now discuss their general features and comment on other approaches.

Our main result is that spacetime noncommutativity has the only effect of deforming the relation between the particle spacetime coordinates $\hat{x}^{\nu}$ and the conjugate variables $\hat{q}^{\nu}$ of the corresponding momenta, replacing the simple identification
\begin{eqnarray}
\hat{x}^{\nu}=\hat{q}^{\nu}
\end{eqnarray}
with the general momentum-dependent formula
\begin{eqnarray}
\hat{x}^{\nu}=\frac{1}{2}\left\lbrace\hat{q}^{\alpha}[\Delta(\ell\hat{p})]_{\alpha}^{\ \nu}+\ell\Sigma^{\nu}(\ell\hat{p})+\mathrm{h.c.}\right\rbrace.
\end{eqnarray}
In particular, the momentum space of our models is not affected by the noncommutativity and the dispersion relation is undeformed. This negative result is quite relevant from a phenomenological point of view, because modified relativistic dispersion relations have been the main target of recent searches for observable quantum gravity effects\cite{quagrap}. It is also at odds with what was proposed for $\kappa$-Minkowski in the pioneering papers\cite{fuzzy1,fuzzy2}, which actually motivated our study. On the positive side, our models exhibit deformed Heisenberg relations, i.e. nontrivial $\Delta_{\mu}^{\ \nu}$, whenever $\Gamma^{\nu\lambda}_{\ \ \alpha}\neq0$. This deformation generally affects the translation properties of both the particle's physical position and the associated quantum uncertainty. We expect the resulting effects to include features of relative locality\cite{relloc} such as those reported in \cite{lateshift,fuzzy2}, which could provide interesting targets for quantum-gravity phenomenology. Even when $\Gamma^{\nu\lambda}_{\ \ \alpha}=0$, the deformed commutation rules among $\hat{m}_{\rho\sigma}$ and $\hat{x}^{\nu}$ should give rise to relative locality effects under Lorentz transformations. Since our framework allows for physical amplitudes to be computed at any order in the deformation parameter $\ell$, this phenomenology can be quantitatively characterized beyond the previous qualitative remarks. We postpone such a detailed analysis to a forthcoming study.

As discussed in the first section, spacetime noncommutativity is usually introduced in a field-theoretical setting, replacing the usual Minkowski background with a noncommutative algebra of coordinates. In this context, continuous groups of kinematical symmetries are described by continuous groups of diffeomorphisms of the background manifold, which are generated by a set of derivation operators acting on the spacetime coordinates. Poincar\'e transformations $(\Lambda,a)$, for example, are described by the background diffeomorphisms
\begin{eqnarray}
x^{\nu}\mapsto\Lambda^{\nu}_{\ \alpha}x^{\alpha}+a^{\nu},
\end{eqnarray}
which are generated by the differential operators
\begin{eqnarray}
P_{\mu}&=&i\hbar\frac{\de}{\de x^{\mu}},\\
M_{\rho\sigma}&=&i\hbar\left(x_{\rho}\frac{\de}{\de x^{\sigma}}-x_{\sigma}\frac{\de}{\de x^{\rho}}\right).
\end{eqnarray}
In order to fully characterize this symmetry group, it is sufficient to specify the action of the generators on the coordinate basis $x^{\nu}$:
\begin{eqnarray}
P_{\mu}\act x^{\nu}&=&i\hbar\delta_{\mu}^{\ \nu},\\
M_{\rho\sigma}\act x^{\nu}&=&i\hbar(x_{\rho}\delta_{\sigma}^{\ \nu}-x_{\sigma}\delta_{\rho}^{\ \nu}),
\end{eqnarray}
and require that 
\begin{eqnarray}
P_{\mu}\act z&=&0,\\
M_{\rho\sigma}\act z&=&0,
\end{eqnarray}
for every $z\in\mathbb{C}$. In fact, the action of the generators on arbitrary analytic functions of $x^{\nu}$ can then be computed by repeated application of the Leibniz rules
\begin{eqnarray}
P_{\mu}\act f(x)g(x)&=&\left[P_{\mu}\act f(x)\right]g(x)+f(x)\left[P_{\mu}\act g(x)\right],\\
M_{\rho\sigma}\act f(x)g(x)&=&\left[M_{\rho\sigma}\act f(x)\right]g(x)+f(x)\left[M_{\rho\sigma}\act g(x)\right].
\end{eqnarray}
This minimal characterization of symmetries is purely algebraic and could in principle make sense even in the noncommutative case, where the standard notion of derivation is not available. However, in order to preserve the nontrivial commutation relations among spacetime coordinates, one cannot proceed as we have done before and deform the action of the generators. In fact, we have proved in our single-particle setting that all possible deformations of this kind must mix particle coordinates and momenta. Since they do not admit a restriction to real space, they cannot be implemented in a field-theoretical context, where there are no particle momenta and the algebra of background coordinates must be mapped onto itself. The only possibility is then to deform the Leibniz rule, turning the ordinary Lie algebra of generators into a Hopf algebra with nontrivial coproduct\cite{hopf}. This modification is the reason why it is necessary to introduce noncommutative transformation parameters\cite{noeth}. The nontrivial coproduct can also affect the commutation properties of symmetry generators and thus lead to deformed relativistic wave equations and dispersion relations. In the special case of $\kappa$-Minkowski, for instance, a popular choice of the coproduct\cite{kappa} induces nonlinear commutation rules among momenta and boost generators and consequently determines a deformed Casimir operator. This description of the $\kappa$-Poincar\'e Hopf algebra is actually the starting point of our main reference \cite{fuzzy1} and the source of our disagreement.

All the difference between our treatment and the usual approach can be traced back to our assumption that deformed Poincar\'e symmetries are standard quantum symmetries described by a Lie algebra of generators. Without this hypothesis, we would have obtained a much wider class of models, including those explored in the seminal works \cite{fuzzy1,fuzzy2}, and we too would probably have relied on Hopf-algebraic considerations to identify the relevant ones. A general classification of this kind is actually available in the literature. In a recent paper\cite{meljanac}, a variety of noncommutative spacetimes as well as the corresponding Hopf symmetries have been characterized in terms of nonlinear realizations of the Heisenberg algebra. Even if the authors adopt a field-theoretical point of view based on Hopf algebras, it is straightforward to recast their results in our single-particle framework and obtain a formal generalization of our models.

The problem with these generalized models is that their symmetry generators do not admit a clear physical interpretation. In fact, nonlinear commutation relations among the generators, as found in \cite{fuzzy1}, cannot hold if the group of finite symmetry transformations is continuous. But if it were discontinuous, then there would be no well-defined infinitesimal transformations to begin with, and symmetry generators would lose their usual physical meaning. By assuming standard quantum symmetries, we found all covariant single-particle models which are not affected by such interpretive difficulties and can directly provide interesting phenomenology. Our approach has the additional advantage of being entirely independent of Hopf-algebraic concepts, thereby avoiding all the problems associated with noncommutative transformation parameters.

\section{Conclusions and outlook}

For various reasons, not least because of the problem of time in standard quantum mechanics, spacetime noncommutativity has so far mostly been studied from a field-theoretical point of view, replacing Minkowski spacetime with some noncommutative algebra of background coordinates. In this paper, inspired by the pioneering works\cite{fuzzy1,fuzzy2}, we advocated and explored the alternative approach of introducing nontrivial commutation relations among particle coordinates into covariant quantum mechanics, providing a full characterization of deformed relativistic symmetries in this framework. In particular, assuming that deformed symmetries admit a standard Lie-algebraic description, we proved that they are just ordinary Poincar\'e transformations acting on some undeformed canonical variables $\hat{p}_{\mu}$ and $\hat{q}^{\nu}(\hat{p}_{\mu},\hat{x}^{\nu})$. This means that momenta retain their usual transformation properties and the hamiltonian constraint, i.e. the relativistic dispersion relation, is undeformed. Spacetime noncommutativity only affects the transformation properties of spacetime coordinates $\hat{x}^{\nu}$ under translations and Lorentz transformations.

A notable feature of our models is the possibility of computing physical amplitudes at any order in the deformation parameter $\ell$. This means that their phenomenology can be quantitatively explored along the lines of \cite{fuzzy2}. We have postponed a detailed phenomenological analysis to a forthcoming study.

In this paper we have considered for simplicity momentum-independent deformation matrices $\Gamma^{\nu\lambda}_{\ \ \alpha}$ and $\Theta^{\nu\lambda}$, but our analysis also applies with little changes to momentum-dependent commutation relations
\begin{eqnarray}
\left[\hat{x}^{\nu},\hat{x}^{\lambda}\right]=\frac{1}{2}i\left(\ell[\Gamma(\ell\hat{p})]^{\nu\lambda}_{\ \ \alpha}\hat{x}^{\alpha}+\ell^{2}[\Theta(\ell\hat{p})]^{\nu\lambda}+\mathrm{h.c.}\right).
\end{eqnarray}
Therefore, our results can be readily extended to include the much-studied Snyder noncommutative spacetime\cite{orisnyd} and similar examples. A further generalization to curved noncommutative spacetimes would be very interesting from a phenomenological point of view, but it is beyond the reach of our framework at the present stage of development. Before attempting to deal with curved noncommutative spacetimes, it is necessary to better understand the physics of covariant quantum mechanics on ordinary curved spacetimes.

Being independent of Hopf-algebraic concepts, our approach could shed some light on the nature of the puzzling noncommutative transformation parameters associated with Hopf symmetry generators. In the simple case of $\kappa$-Minkowski, it is actually possible to express these objects in terms of single-particle momenta\cite{fuzzy1}. Our framework could be used to find such a representation for generic spacetime noncommutativity.

\vfill
The author declares that there is no conflict of interest regarding the publication of this paper.

\end{document}